\documentclass[publi-ica4,french]{IC2015}

\usepackage{verbatim}

\usepackage[utf8]{inputenc}
\usepackage[T1]{fontenc}
\usepackage{xcolor}
\usepackage[backref=page,hidelinks]{hyperref}
\usepackage{amsmath}
\usepackage{amssymb}
\usepackage{mathrsfs}
\usepackage{mathtools}
\usepackage{mathabx}
\usepackage{dsfont}
\usepackage{graphicx}
\usepackage{float}
\usepackage{adjustbox}
\usepackage{multirow}

\providecommand{\keywords}[1]{\textbf{\textit{Keywords---}} #1}

\providecommand{\example}[1]{\textit{#1}}

\shortouvrage{}

\title{Enhancing Translation Language Models with Word Embedding for Information Retrieval}

\author{Jibril Frej, Jean-Pierre Chevallet, Didier Schwab }

\institute{
 \textsc{Laboratoire d’informatique de Grenoble (LIG)\\
Univ. Grenoble Alpes (UGA)}\\ 
\texttt{jibril.frej@etu.univ-grenoble-alpes.fr} \\
\texttt{jean-pierre.chevallet@imag.fr}\\
\texttt{didier.schwab@imag.fr}
}

\begin{document}
\maketitle

\begin{abstract}
  In this paper, we explore the usage of Word Embedding semantic resources for Information Retrieval (IR) task. This embedding, produced by a shallow neural network, have been shown to catch semantic similarities between words \citep{article6}. Hence, our goal is to enhance IR Language Models by addressing the term mismatch problem. To do so, we applied the model presented in the paper \textit{Integrating and Evaluating Neural Word Embedding in Information Retrieval} by \cite{article2} that proposes to estimate the translation probability of a Translation Language Model using the cosine similarity between Word Embedding. The results we obtained so far did not show a statistically significant improvement compared to classical Language Model.

\end{abstract}

%
%
\section{Introduction}

Information Retrieval Systems (IRS) are computer assistants that help to retrieve digital documents, in which user is supposed to found relevant information for his task. Hence an IRS \emph{is} about semantics, because user information needs is topically related and serve to help to accomplish user's task.

\vspace{0.3em}

Curiously, most commercial and experimental search engine do not handle any sort of semantics nor knowledge to solve queries. This is because matching computations are  mostly based on statistical word distributions, and intersection between query and documents. Also, most IR models see query and document as simple bag of word. 
Though, these systems provide satisfaction to their user, as long as statistics are possible, i.e. documents are long enough and queries are expressed using the same vocabulary as documents. 

\vspace{0.3em}

When we are not in this situation, that is, if documents to be retrieved are very short and/or there is a strong discrepancy between document vocabulary and user's one, IRS are facing the problem of \emph{term mismatch}. For example, the collection Europeanna\footnote{http://www.europeana.eu}  gives access to millions of digital objects from cultural heritage, by a very small textual meta-data description. Even, descriptions are made by specialists that are prone to use technical terms.

\vspace{0.3em}

Hence, in this paper we propose to study the effect of exploiting semantic resources to reduce term mismatch negative effect on collection with specialized vocabularies. We focus on automatically constructed resources, namely word Embedding resources, first because they cover a very large vocabulary, and second, because they seem to capture interesting word semantics \citep{article6}.

\vspace{0.3em}

A very simple way to exploit a resource to solve term mismatch, is to expand queries or documents with words that are semantically similar according to the resource. This approach has been heavily studied \citep{Carpineto:2012:SAQ:2071389.2071390} and suffers from a definitive problem: how to control query meaning shift when a-priory the query is statically modified, i.e. for all documents ?

\vspace{0.3em}

In this paper, we propose not to change the query nor the document but to adapt the matching function. This adaptation requires to change the formula that computes the Relevant Status Value (RSV). This formula depends on the Information Retrieval (IR) model. 

\vspace{0.3em}

There are several large categories of IR model: Vector Space, Logical Based, Probabilist, Graph Based, and Language Models (LM). Among these models, only the Probabilist models like the famous BM25 formula of Robertson \citep{robertson1995okapi}, or Language Models provide state of the art results. Some recent proposals of IR Graph Modeling \citep{DBLP:conf/f-ic/BannourZN16} have the advantage to fuse one single model document index and knowledge base, and exploit only one matching function: activation propagation from query term to document through indexing terms and knowledge concept nodes. Beside the nice property of this model to exploit a Knowledge Base at query time, without query or document expansion, experiments of this model are still bellow the state of the art of Probabilistic or Language models. For this reason, we have decided to work on the transformation of a Language Model formula. We did not chose a Probabilistic model because formulas of the LM model are much simpler for similar results \citep{introIR}, hence they are easier to transform. 

\vspace{0.3em}


A simple way to include a Knowledge Resource in a LM matching function, is  Translation Language Models \citep{Stat_trans} where the translation probability between query and document terms is taken into account, in addition to the exact term matching. Probability are estimated using the mutual information between the two terms and is the highest if the considered terms have the same distribution over the collection. This computation can be considered as an automatic basic knowledge resource extracted from the corpus itself. Recently, it has been proposed to estimate the translation probability using Neural Word Embedding \citep{article2}.

\vspace{0.3em}

Word Embedding denotes a set of methods to produce Knowledge resources with vector representation of words\footnote{Most of the time the vectors are real-valued.} that express some semantics learned from word usage in very large text collection. Usually these methods are based on the distributional hypothesis \citep{harris54}: \emph{words that occur in the same contexts tend to have similar meaning}. The vector representation of words are computed using their context so that words with similar meaning will have similar vector representation. Word Embedding includes dimension reduction techniques on the word co-occurrence matrix (Latent Semantic Analysis \citep{LSA}), probabilistic models (Latent Dirichlet Allocation \citep{LDA}) and more recently shallow neural network-based methods such as the skip-gram model that can also learn phrases vector representations and is very effective on word similarity and word analogy tasks \citep{article7}. As we said before, these vector representations can be used to capture semantic relationships between words by measuring the similarity between the vectors, with the cosine similarity measure for example. 

\vspace{0.3em}

Such vector representation could help us with the vocabulary mismatch problem since they provide us  a new way to estimate the translation probability between query and document terms by considering their semantic similarity. The empirical results show that improving Dirichlet Language Model using Word Embedding is possible.

\vspace{0.3em}

In the rest of this paper, we first recall LM formulas in section \ref{LM},  present the extension with the translation model in section \ref{TLM}, then the usage of Word Embedding within this model in section \ref{WE}. We present the implementation of this model in section \ref{data} and results on the Cultural Heritage in CLEF (CHiC) Dataset\footnote{http://ims.dei.unipd.it/data/chic/}, a sub part of the Europeanna, in section \ref{Results}.

%

\section{Language models}
\label{LM}

In this part, we recall basics on the Language Model in order to detail our modifications of the formula in the next section. The query likelihood Language Model aims to  rank  documents $d$ by computing the probability of a document  interpreted as the likelihood that it is relevant to the query $q$. Hence, the Relevance Status Value (RSV) of a Language Model based IR is expressed by:
 \begin{equation}
 RSV(q,d) = p(d|q)
 \end{equation}
 
 Using Bayes rule, we have: 
 
 \begin{equation}
 RSV(q,d) = \frac{p(q|d) p(d)}{p(q)}
 \end{equation}
 
 We can ignore the constant $p(q)$ to rank documents and we also consider that $p(d)$ is uniform over all the documents of the collection and can also be ignored. Therefore, to rank documents with respect to a given query, we only have to compute the probability of having $q$, knowing $d$. In LM model, document $d$ is replaced by its language model, i.e. the probability distribution of terms of the vocabulary in $d$ denoted $\theta_d$. So the RSV is:
 
 \begin{equation}
 RSV(q,d) \simeq_{rank} p(q|\theta_d)
 \end{equation}

 In this work, we chose the multinomial event model for $\theta_d$, and the unigram language model\footnote{Usually in IR the unigram language model give the best results with the lowest computational cost \citep{introIR}}, so the probability of a given word does not depend on its context: 

\begin{equation}
   RSV(q,d) \simeq_{rank} log \left( p(q|\theta_d) \right) = \sum\limits_{i = 1 }^{|q|} log\left( p(q_i|\theta_d) \right)   
    \label{eq1}
\end{equation}

With $q_i$ the $i^{th}$ term of the query and $|q|$ the query size. To estimate $p(q_i|\theta_d)$, Dirichlet Language Model is used since it has been shown to work better with Translation Language Models \citep{article4} :   

\begin{equation}
    p(q_i|\theta_d) = \frac{|d|}{\mu + |d|} p_{ml}(q_i | \theta_d) + \frac{\mu}{\mu + |d|}p(q_i|C) 
    \label{eq2}
\end{equation}

$\mu \in \mathds{R}^+$ is the smoothing parameter and $p_{ml}(q_i | \theta_d)$ is estimated using the maximum likelihood \footnote{We consider that terms in a document follow a multinomial distribution} and is equal to:
\begin{equation}
p_{ml}(q_i | \theta_d) = \frac{c(q_i , d)}{|d|} 
\end{equation}
With $c(q_i , d)$ the frequency of $q_i$ in $d$ and $|d|$ the document size. The same goes for $p(q_i|C)$ the smoothing term which is also estimated with the maximum likelihood : $p(q_i|C) = c(q_i , C)/|C|$. This estimation of $p(q_i|\theta_d)$ leads to the following ranking formula (see \hyperref[demo1]{\textit{the appendix}} for more details) :

\begin{equation}
    RSV(q,d) \simeq_{rank}  \sum\limits_{i : c(q_i , d) > 0} \left[ log\left( 1 + \frac{c(q_i , d)}{\mu p(q_i | C)} \right) \right] + |q|log \left(\frac{\mu}{\mu + |d|} \right)
    \label{eq3}
\end{equation}

We decided to present equations (\ref{eq2}) and (\ref{eq3}) even though they are equivalent for ranking documents because both of them can be found to describe Dirichlet Language Model. In the frame of our work, and for reasons that are developed in \hyperref[para]{\textit{the appendix}}, we will introduce the next models by giving the expression of $p(q_i|\theta_d)$ as in equation (\ref{eq2}). 

\vspace{0.3em}

As we said previously, one issue of these models is the term mismatch problem: as we can see on equation (\ref{eq3}) ranking takes into account only the terms that appear in both the considered document and query. Consequently, relevant documents that do not contain the exact query terms will not be considered. One approach to solve this problem is to adapt the language model to take into account the semantic similarities between terms.

\section{Translation Language Models}
\label{TLM}
Translation Language Models (TLM) try to estimate the semantic similarity between two terms by using tools from statistical translation. The main idea is to estimate the likelihood of translating a document to a query using the translation probability between terms \citep{article4}. To do so, the maximum likelihood estimator in the Dirichlet language model $p_{ml}(q_i | \theta_d)$ is replaced with the likelihood that the query has been produced by a translation of the document $ p_t(q_i|\theta_d)$:

\begin{equation}
    p(q_i|\theta_d)  =   \frac{|d|}{\mu + |d|} p_t(q_i|\theta_d) + \frac{\mu}{\mu + |d|}p(q_i|C) 
\end{equation}
\noindent
$p_t(q_i|\theta_d)$ is calculated the following way :  

\begin{equation}
    p_t(q_i|\theta_d) = \sum_{u \in d} p_t(q_i|u)p_{ml}(u|\theta_d) 
\end{equation}

With $p_t(w|u)$ the probability to translate term $u$ into term $w$ which is estimated using mutual information between $u$ and $w$ \citep{article4}  :

\begin{equation}
    p_t(w|u) = \frac{I(w , u)}{ \sum\limits_{w' \in V}I(w' , u) }
\end{equation}

$I(w,u)$ is the mutual information score between word $u$ and $w$, defined as follow :

\begin{equation}
    I(w,u) = \sum\limits_{X_w = 0,1} \sum\limits_{X_u = 0,1} p(X_w , X_u) log \left( \frac{p(X_w , X_u)}{p(X_w) p( X_u)} \right)
\end{equation}

With $X_w$ and $X_u$ binary random variables indicating if a word is absent or present (refer to \citep{article4} for more details). 

\section{Word Embedding-based Translation Language Model }
\label{WE}
Within the frame of this work, word Embedding are used instead of mutual information in order to estimate the translation probability $p_t(q_i|u)$. The model is named Word Embedding-based Translation Language Model (WETLM). We consider the new estimation of the translation probability, denoted $p_{cos}(q_i|u)$,  to be proportional to the similarity between $q_i$ and $u$ that is measured with the cosine between the vectors of the two terms : 

\begin{equation}
    p_{cos}(q_i|u) = \frac{cos(q_i , u)}{\sum\limits_{u' \in V} cos(u' , u) }
    \label{trans_proba}
\end{equation}

Consequently the ranking formula becomes : 

\begin{equation}
    p(q_i|\theta_d)  =  \frac{|d|}{\mu + |d|} p_{cos}(q_i|\theta_d) + \frac{\mu}{\mu + |d|}p(q_i|C)
    \label{rank_formula}
\end{equation}

With $p_{cos}(q_i|\theta_d) = \sum_{u \in d} p_{cos}(q_i|u) p_{ml}(u|\theta_d)$. Both the estimation of the translation probability $p_t$ and $p_{cos}$ underestimate the self-translation probability: we can have $p_{cos}(u|w) > p_{cos}(u|u)$ for $u \neq w$ which is not desirable for a translation language model \citep{axiom}. 

One way to make sure that the self-translation probability is the highest for a given term is to redefine it by introducing a hyper-parameter $\alpha \in [0,1]$ that "controls" the self-translation probability \citep{article4}:

\begin{equation}
p_{cos-\alpha}(w|u) = \begin{cases} \alpha + (1 - \alpha)p_{cos}(w|u) & \mbox{if } u=w \\ (1 - \alpha)p_{cos}(w|u) & \mbox{if } u \neq w \end{cases}
\end{equation}

This formula ensures the fact that we have $p_{cos-\alpha}(u|u) > p_{cos-\alpha}(u|w)  \forall u , w \in V$ for $\alpha > 0.5$. The model that uses $p_{cos-\alpha}$ to estimate the translation probability will be referred as WETLM-$\alpha$. We set the value of $\alpha$ to 0.45 since it is the one that produced the best results for the Threshold T = 0.7. This approach was not developed in the paper \textit{Integrating and Evaluating Neural Word Embedding in Information Retrieval} by \cite{article2} since they reported that the word Embedding they used did not underestimate the self-translation probability.

\section{Implementation and data}
\label{data}

For this work, instead of using an already existing Information Retrieval System (IRS) such as \href{http://terrier.org/}{\textit{Terrier}}, we developed our own IRS in C++ to easily add word Embedding to the classical models and also because having a low retrieval time\footnote{Which is the case with Terrier} is not an objective: before trying to compute a fast IRS, we should make sure that the Word Embedding-based language models outperforms state of the art Language models. In order for our results to be comparable with other work that used Terrier, we made sure that with the same pre-processing on the corpus, we obtain the same results as Terrier ( see \hyperref[para3]{\textit{the appendix}} for more details about our IRS  ). When doing so, we noticed that Terrier does not implement Dirichlet language model using equation (\ref{eq3}) (more details in \hyperref[para2]{\textit{the appendix}}).

\vspace{0.3em}

During pre-processing, we used a Stop List and replaced capital letters with lower case letters on both the collection and the queries. In order to have the same results as Terrier with our IRS, we also removed characters that were not digits or letters and deleted words that contained more that 4 digits or more than 3 consecutive identical characters. We did not use any stemmer on our collection since the best results were obtained without any stemming.

\vspace{0.3em}

We used the Cultural Heritage in CLEF (CHiC) 2012 English collection for ad hoc retrieval. This collection is composed of 1 107 176 documents containing "metadata  records describing digital representations of cultural  heritage objects" \citep{article5} and 50 queries for \textit{ad hoc} retrieval tasks. Below is a table summing up some statistics of the collection:

\begin{table}[ht!]
    \centering
    \begin{tabular}{|c|c|c|c|c|}
    \hline
     \#d & Avdl & Vocabulary Size & \#q & Avql \\
    \hline
    1 107 176 & 30.92 & 290 265 & 50 & 1.84\\
    \hline
\end{tabular}
    \caption{CHiC 2012 statistics }
\end{table}

With $\#d$ and $\#q$ being respectively the number of documents in the collection and the number of queries. Avdl is the average document length and Avql is the average query length. To evaluate the models, the top 1000 documents were returned for each query, the MAP and P@10 were computed using the standard tool for evaluating an \textit{ad hoc} retrieval: \href{http://trec.nist.gov/trec_eval/}{\textit{trec\_eval}}. 

\vspace{0.3em}

We used the word2vec-GoogleNews-vectors word Embedding of dimension 300, pre-trained on the google news Corpus (3 billion words) that are available \textit{here} \footnote{\url{https://github.com/mmihaltz/word2vec-GoogleNews-vectors}}. As we said earlier, the CHiC collection we used have a very specific vocabulary and even if the word Embedding we used were trained on a 3 billion words corpus, a lot of word of the vocabulary were missing :

\begin{itemize}
    \item only 42.68\% of the words of the vocabulary have an Embedding;
    \item but 91.92\% of word occurrences in the collection have an Embedding.
\end{itemize}

On the other hand, most of the queries's terms had an Embedding : 

\begin{itemize}
    \item 94.95\% of the queries terms have an Embedding;
    \item 2\% of the queries have none of their term that posses an Embedding.
\end{itemize}

\vspace{0.3em}

Finally, the translation probability described in equation (\ref{trans_proba}) is not the one that was implemented: we computed the cosine similarity between two words if it was above a given threshold \textbf{T}. This allowed us to reduce the number of cosine similarities to compute and also it acts like a noise reducer since we did not take into account the similarity between non similar terms. We found that \textbf{T} = 0.7 produces the highest MAP on the CHiC collection using word2vec-GoogleNews-vectors.

\section{Results}
\label{Results}

Instead of giving the value of the parameter $\mu$ that produces the optimal results, we decided to display the MAP for a range of values of $\mu$ to see if the WETLM outperforms (or not) the Dirichlet LM consistently or if for some values of $\mu$ one model performs better than the other.

\vspace{0.3em}

At first we evaluated the Dirichlet LM, the optimal value we found for $\mu$ was 44. Retrieved documents are evaluated using the Mean Average Precision and P@10: 

\begin{table}[H]
    \centering
    \begin{tabular}{|c|c|c|c|c|c|c|c|c|c|c|}
    \hline
     $\mu$ & 12 & 16 & 20 & 24 & 28 & 32 & 36 & 40 & 44 & 48\\
    \hline
    MAP (\%) & 35.61 & 36.09 & 36.16 & 36.24 & 36.30 & 36.36 & 36.39 & 36.23 & 36.43 & 36.09\\
    \hline
    \hline
    $\mu$ & 52 & 56 & 60 & 64 & 68 & 72 & 76 & 80 & 84 & 88\\
    \hline
     MAP (\%) & 36.05 & 36.06 & 35.82 & 35.86 & 35.92 & 35.92 & 36.04 & 35.77 & 35.68 & 35.68\\
    \hline
\end{tabular}
    \caption{Values of the MAP on the CHIC2012 collection using the Dirichlet Language Model}
    \label{dirichlettable}
\end{table}

\begin{table}[H]
    \centering
    \begin{tabular}{|c|c|c|c|c|c|c|c|c|c|c|}
    \hline
     $\mu$ & 12 & 16 & 20 & 24 & 28 & 32 & 36 & 40 & 44 & 48\\
    \hline
    P@10 (\%) & 33.54 & 33.75 & 33.75 & 33.96 & 34.17 & 34.38 & 34.17 & 34.17 & 34.38 & 34.38\\
    \hline
    \hline
    $\mu$ & 52 & 56 & 60 & 64 & 68 & 72 & 76 & 80 & 84 & 88\\
    \hline
     P@10 (\%) & 34.38 & 34.17 & 34.38 & 34.38 & 34.38 & 34.38 & 34.38 & 34.38 & 34.38 & 34.58\\
    \hline
\end{tabular}
    \caption{Values of the P@10 on the CHIC2012 collection using the Dirichlet Language Model}
    \label{dirichlettableP10}
\end{table}

We checked that the results obtained are identical to the ones produced by Terrier with the same pre-processing on the collection. Also we decided to explore values of $\mu$ that are close to the Average Document Length (Avdl) since the optimal value of $\mu$ in the Dirichlet Language Model is usually around the Avdl. 
Table \ref{Embeddingtable} below represents the results obtained with the WE-based Translation Language Model :

\begin{table}[H]
    \centering
    \begin{tabular}{|c|c|c|c|c|c|c|c|c|c|c|}
    \hline
     $\mu$ & 12 & 16 & 20 & 24 & 28 & 32 & 36 & 40 & 44 & 48\\
    \hline
    MAP (\%) & 36.89 & 37.71 & 37.76 & 37.86 & 37.78 & 37.79 & 37.81 & 37.65 & 37.67 & 37.29\\
    \hline
    \hline
    $\mu$ & 52 & 56 & 60 & 64 & 68 & 72 & 76 & 80 & 84 & 88\\
    \hline
     MAP (\%) & 37.17 & 36.87 & 36.66 & 36.69 & 36.57 & 36.56 & 36.55 & 36.50 & 36.37 & 36.35\\
    \hline
\end{tabular}
    \caption{Values of the MAP on the CHIC2012 collection using WETLM}
    \label{Embeddingtable}
\end{table}

\begin{table}[H]
    \centering
    \begin{tabular}{|c|c|c|c|c|c|c|c|c|c|c|}
    \hline
     $\mu$ & 12 & 16 & 20 & 24 & 28 & 32 & 36 & 40 & 44 & 48\\
    \hline
    P@10 (\%) & 34.38 & 34.38 & 35.21 & 35.42 & 35.63 & 35.63 & 35.42 & 35.42 & 35.42 & 35.63\\
    \hline
    \hline
    $\mu$ & 52 & 56 & 60 & 64 & 68 & 72 & 76 & 80 & 84 & 88\\
    \hline
     P@10 (\%) & 35.42 & 35.21 & 35.21 & 35.21 & 35.21 & 35.21 & 35.21 & 35.21 & 35.21 & 35.21\\
    \hline
\end{tabular}
    \caption{Values of the P@10 on the CHIC2012 collection using WETLM}
    \label{EmbeddingtableP10}
\end{table}

As we can see the WE-based Translation Language Model seems to sightly outperform the Dirichlet Language model for every $\mu$. The optimal value of $\mu$ we found is different for the two models : $\mu_{opt} = 44$ for the Dirichlet LM and $\mu_{opt} = 24$ for the WETLM. We performed a paired t-test over the average precision of each query for $\mu$ = 36 for both models, the measured p-value with R is $0.1733$ > 0.01. Unfortunately the improvement is not statistically significant. 

\vspace{0.3em}

In the table below we show some of the results collections presented in the work of \citep{article2} over the 3 collections AP88-89 , WSJ87-92 and  DOTGOV : 

\begin{table}[H]
    \centering
    \begin{tabular}{|c|c|c|c|c|c|c|}
    \hline
    \multirow{2}{*}{Method} &  \multicolumn{2}{c|}{AP88-89 } & \multicolumn{2}{c|}{WSJ87-92} &  \multicolumn{2}{c|}{DOTGOV}  \\
    \cline{2-7}
     & MAP & P@10 & MAP & P@10 & MAP & P@10  \\
    \hline
    Dirichlet LM & 22.69 & 39.60 & 21.71 & 40.80 & 18.73 & 24.60 \\
    \hline
    WETLM & 24.27\textbf{*} & 41.00 & 22.66\textbf{*} & 42.40\textbf{*} & 19.32 & 25.00  \\
    \hline
\end{tabular}
    \caption{Values of the MAP and P@10 reported by \citep{article2} on the collections AP88-89 , WSJ87-92 and  DOTGOV using Dirichlet LM and WETLM. The statistically significant differences are indicated by \textbf{*}. }
    \label{Zucconres}
\end{table}

As we can see, according to \citep{article2}, depending on the collection, the WETLM can produce results that exhibit a statistically significant improvement of the MAP compared to Dirichlet LM. 

\vspace{0.3em}

Table \ref{Embeddingalphatable} below represents the results obtained with the WE-based Translation Language Model that "controls" the self translation probability with the parameter $\alpha$:

\begin{table}[H]
    \centering
    \begin{tabular}{|c|c|c|c|c|c|c|c|c|c|c|}
    \hline
     $\mu$ & 12 & 16 & 20 & 24 & 28 & 32 & 36 & 40 & 44 & 48\\
    \hline
    MAP (\%) & 37.35 & 37.92 & 38.07 & 38.15 & 38.28 & 38.31 & 38.35 & 38.19 & 38.17 & 37.81\\
    \hline
    \hline
    $\mu$ & 52 & 56 & 60 & 64 & 68 & 72 & 76 & 80 & 84 & 88\\
    \hline
     MAP (\%) & 37.72 & 37.73 & 37.53 & 37.58 & 37.59 & 37.47 & 37.40 & 37.10 & 37.06 & 37.05\\
    \hline
\end{tabular}
    \caption{Values of the MAP on the CHIC2012 collection using WETLM-$\alpha$}
    \label{Embeddingalphatable}
\end{table}

\begin{table}[H]
    \centering
    \begin{tabular}{|c|c|c|c|c|c|c|c|c|c|c|}
    \hline
     $\mu$ & 12 & 16 & 20 & 24 & 28 & 32 & 36 & 40 & 44 & 48\\
    \hline
    P@10 (\%) & 34.79 & 34.79 & 35.21 & 36.04 & 36.25 & 36.46 & 36.46 & 36.46 & 36.25 & 36.67\\
    \hline
    \hline
    $\mu$ & 52 & 56 & 60 & 64 & 68 & 72 & 76 & 80 & 84 & 88\\
    \hline
     P@10 (\%) & 36.25 & 36.25 & 36.25 & 36.25 & 36.25 & 36.25 & 36.25 & 36.25 & 36.25 & 36.25\\
    \hline
\end{tabular}
    \caption{Values of the P@10 on the CHIC2012 collection using WETLM-$\alpha$}
    \label{EmbeddingalphatableP10}
\end{table}

We performed a paired t-test over the average precision of each query for $\mu$ = 36 to compare LM and WETLM-$\alpha$ models : the measured p-value with R is $0.01219$ > 0.01 : the improvement is still not statistically significant.

\section{Conclusion and future work}

The results we obtained are consistent with the ones in \citep{article2} since they observed the same improvement as we did in the MAP. They also checked that their improvements were independent of the corpus and the training set for the Word Embedding: they do not need to be trained on the same corpus used in retrieval. For now our results are limited to one corpus and one set of word embedding, one of our objective in the near future is to perform the experiments on different corpora and also to improve our model by considering the context of the terms of the query by using phrase vectors \citep{article7} to replace the query or to perform query expansion and by modifying the translation probability so that it satisfies a set of constraints \citep{axiom}.

\bibliography{IC2015}

\newpage

\section{Appendix}

\subsection{From equation (\ref{eq2}) to equation (\ref{eq3}) :}

\begin{equation*}
    \begin{split}
     &log \left( p(q|\theta_d) \right) = \sum\limits_{i = 1 }^{|q|} log\left( p(q_i|\theta_d) \right)\\  
     =& \sum\limits_{i = 1 }^{|q|} log\left( \frac{|d|}{\mu + |d|} p_{ml}(q_i | \theta_d) + \frac{\mu}{\mu + |d|}p(q_i|C) \right)\\ 
     =& \sum\limits_{i = 1 }^{|q|} log\left( \frac{c(q_i , d) + \mu p(q_i | C)}{|d| + \mu} \right)\\ 
     =& \sum\limits_{i : c(q_i , d) > 0} log\left( \frac{c(q_i , \theta_d) + \mu p(q_i | C)}{|d| + \mu} \right) + \sum\limits_{i : c(q_i , d) = 0} log\left( \frac{ \mu p(q_i | C)}{|d| + \mu} \right)\\
     =& \sum\limits_{i : c(q_i , d) > 0} log\left( \frac{c(q_i , d)  + \mu p(q_i | C)}{|d| + \mu} \right) - \sum\limits_{i : c(q_i , d) > 0} log\left( \frac{ \mu p(q_i | C)}{|d| + \mu} \right) +  \sum\limits_{i = 1}^{|q|} log\left( \frac{ \mu p(q_i | C)}{|d| + \mu} \right) \\
     =& \sum\limits_{i : c(q_i , d) > 0} log\left( \frac{c(q_i , d) + \mu p(q_i | C)}{|d| + \mu} \times \frac{|d| + \mu}{\mu p(q_i|C)} \right) + \sum\limits_{i = 1}^{|q|} log\left( \frac{ \mu p(q_i | C)}{|d| + \mu} \right)\\
     =& \sum\limits_{i : c(q_i , d) > 0} log\left( 1 + \frac{c(q_i , d) }{ \mu p(q_i | C)} \right) + |q| log \left( \frac{\mu}{|d| + \mu}\right) +  \sum\limits_{i = 1}^{|q|} log\left( p(q_i | C) \right)\\
     \end{split}
    \label{demo1}
\end{equation*}

The last term of the equation above depends only on the query and collection, therefore it  can be ignored to rank documents, which leads to :

\begin{equation*} 
     log \left( p(q|\theta_d) \right) = \sum\limits_{i : c(q_i , d) > 0} \left[ log\left( 1 + \frac{c(q_i , d)}{\mu p(q_i | C)} \right) \right] + |q|log \left(\frac{\mu}{\mu + |d|} \right)
\end{equation*}

\subsection{Why equation (\ref{eq2}) instead of equation (\ref{eq3}) :}\label{para}

\vspace{0.5em}

As we said previously, usually language model RSV functions are presented using a sum that goes through terms that appear in both the document and the query, as in  equation (\ref{eq3}). The reason is that this formulation is more compatible with the usage of an inverted index. In fact an inverted index just associates to a term, all documents with non null term frequency: this is called the posting list. When using inverted index, the matching computation have to compute partial RSV sum for all documents found in posting lists, in parallel. If the formula only requires terms to have non null frequency in documents as in formula (\ref{eq3}) the matching algorithm is trivial. If the formula need all terms of the query, as in formula (\ref{eq2}) then some partial matching sums has to be corrected during matching computation and the matching algorithm is much more complex and less efficient.

In our experimental system, we do not use any inverted index: we store the all index in a direct structure in main memory. Hence,  matching is done by simply browsing all documents and computing RSV strait using the formula.

The models we presented that used statistical translation and word Embedding cannot be written with only a sum over $q \cap d$ as in equation (\ref{eq3}) : this is why we used equation (\ref{eq2}) to give the expression of the rank. Hence the implementation using a inverted fil is more problematic, and for example, cannot be done in a simple way in Terrier. That is the reason why we could not use Terrier for these experiments.

\subsection{Terrier's Dirichlet Language Model :}\label{para2}

\vspace{0.5em}

Moreover during the experiment, when we tried to obtain the same results as Terrier, we noticed that they did not implement exactly equation (\ref{eq2}) to compute the score of each document. Below is the formula they used to compute the score of documents : 

\begin{equation}
    log \left( p(q|\theta_d) \right) = \sum\limits_{i : c(q_i , d) > 0} \left[ log\left( 1 + \frac{c(q_i , d)}{\mu p(q_i | C)} \right) + log \left(\frac{\mu}{\mu + |d|} \right) \right]
    \label{terriereq4}
\end{equation}

Compared to equation (\ref{eq2}) where the quantity $log \left(\frac{\mu}{\mu + |d|} \right)$ is computed |q| times, this formula computes it only $|q \cap d|$ times. Since $log \left(\frac{\mu}{\mu + |d|} \right) < 0 $, equation(\ref{terriereq4}) gives a little bit more importance to query terms that do not appear in documents. For our experiment, the Dirichlet language model was implemented using equation (\ref{eq2}) but we also checked that we obtained the same results as Terrier when computing equation (\ref{terriereq4}).

\vspace{1em}

\subsection{Some details about the implemented Information Retrieval System }\label{para3}

\vspace{0.5em}

In order to have the same results as Terrier, our IRS performed the following pre-processing on the collection :

\begin{itemize}
    \item We replaced with a space all the characters that were not integers or letters, for example the term \example{pre-processing} is broken down into the two terms \example{pre} and \example{processing}.
    \item We also replaced majuscule letters with their minuscule equivalent
    \item We deleted terms that contained more than 4 digits
    \item We deleted terms that had more than 3 consecutive identical characters 
\end{itemize}

\vspace{0.3em}

Also, when we implemented equation (\ref{rank_formula}) ( the ranking formula of the WETLM), we decided to remove the normalization terms when  $p_{cos}(q_i , \theta_d) $ or $p(q_i|C)$ were equal to 0 :

\begin{equation}
p(q_i|\theta_d)  = 
\begin{cases}   \vspace{0.5em}
\frac{|d|}{\mu + |d|} p_{cos}(q_i|\theta_d) + \frac{\mu}{\mu + |d|}p(q_i|C)  & \mbox{if } p_{cos}(q_i|\theta_d) \neq 0 \mbox{ and } p(q_i|C) \neq 0  \\ \vspace{0.5em}
 p_{cos}(q_i|\theta_d)   & \mbox{if } p_{cos}(q_i|\theta_d) \neq 0 \mbox{ and } p(q_i|C) = 0 \\ \vspace{0.5em}
p(q_i|C)  & \mbox{if } p_{cos}(q_i|\theta_d) = 0 \mbox{ and } p(q_i|C) \neq 0 
\end{cases}
\end{equation}

We decided to do so in order to avoid underweighting the translation probability of a query term that is not in the collection and conversely to avoid underweighting the smoothing term of associated to a word that is in the collection but does not have any similar terms in the considered document.

\end{document}